\documentclass[twocolumn,10pt]{tsfp}
\usepackage{flushend}
\usepackage{graphicx}
\usepackage[authoryear,round]{natbib}
\usepackage{fancyhdr}
\usepackage{physics}
\usepackage{mathtools}
\DeclarePairedDelimiter{\avg}{\langle}{\rangle}
\usepackage[font=small]{caption}
\usepackage{subcaption}
\usepackage{tikz}
\usepackage{hyperref}
 
\title{TOWARDS PR-DNS OF SCOUR AROUND A WALL-MOUNTED CYLINDER IN TURBULENT OPEN CHANNEL FLOW}

\author{Leo Bürk
    \affiliation{
	Institute for Water and Environment\\
	Karlsruhe Institute of Technology\\
	76131 Karlsruhe, Germany\\
    leo.buerk@kit.edu
    }	
}

\author{Artjom Hermann
    \affiliation{
    PS-LE/EIS2\\
	Robert Bosch AG\\
	5400 Hallein, Austria\\
	artjom.hermann@at.bosch.com
    }
}

\author{Markus Weyrauch
    \affiliation{
    Center of Applied Space Technology and Microgravity (ZARM)\\
	University of Bremen\\
	28359 Bremen, Germany\\
	markus.weyrauch@zarm.uni-bremen.de
    }
}

\author{Markus Uhlmann
    \affiliation{
    Institute for Water and Environment\\
	Karlsruhe Institute of Technology\\
	76131 Karlsruhe, Germany\\
	markus.uhlmann@kit.edu
    }
}

\begin{document}

\maketitle   
\thispagestyle{fancy}

\fontsize{9}{11}\selectfont

\section*{ABSTRACT}

Particle-resolved direct numerical simulation (PR-DNS) is performed for turbulent open channel flow over a smooth horizontal wall with a vertical cylinder and a dilute set of mobile, heavy, spherical particles. At the chosen parameter point (which matches a previous study without a cylinder) the particles are mostly translating in the horizontal plane while remaining in contact with the wall. It is shown that the presence of the cylinder leads to the generation of intense vortical structures, enhanced turbulence intensity in the wake region, and to strong modifications of the local wall shear stress. These cylinder-induced perturbations have direct consequences for the average particle concentration: preferential accumulation/depletion in different parts of the wake region occurs, while the wall-normal transport of particles (against gravity) is significantly enhanced. A second simulation which adds roughness elements on the wall reveals an additional effect upon the wall-normal distribution of particles. It turns out that the configuration with wall-roughness and a wall-mounted cylinder features the largest fraction of entrained particles, even far from the wall.

\section*{Introduction}

Scour refers to the erosion of sediment material due to the local flow conditions in the vicinity of a hydraulic structure. In practical applications this tends to weaken the foundations which in turn can ultimately lead to structural failure \cite[]{melville2000}. Despite the obvious interest in predicting the occurrence of scour and of its extent, today’s numerical modeling approaches still exhibit significant uncertainties. In purely Eulerian formulations the sediment phase is treated as a continuum, either with a two-fluid approach \cite[]{nagel2020} or in terms of a sediment continuity equation \cite[]{khosronejad2012}. In the predominant Eulerian-Lagrangian approach the sediment is described as (unresolved) point-particles \cite[]{li2018}. All of the above techniques require closure models for fluid-particle exchange terms which arise due to averaging. In the PR-DNS approach \cite[e.g.][]{uhlmann2023}, on the other hand, no such models are needed, since the flow around each particle is resolved numerically, thereby yielding all ingredients to drive the coupled system. While this technique is now well established in the particulate flow community, it has previously not been used to tackle the problem of scour.

In the present contribution we report on the first steps towards performing PR-DNS of scour around a wall-mounted cylinder. For this purpose we first omit the thick sediment bed and instead focus on the motion of a limited number of mobile, heavy particles in the flow over a smooth wall. Parameter values are chosen as close as possible to the previous investigation of \cite{kidanemariam2013direct} which does not feature a cylindrical obstacle. This step allows us to investigate the influence of the cylinder upon the boundary layer with emphasis on the consequences for the particle motion. Secondly, a fixed layer of particles is included at the wall to address the influence of roughness on the particle motion. \\
Some of the questions we are attempting to address with the present contribution read as follows: Do the particles avoid the cylinder wake? Does the flow modification induced by the cylinder have an impact on the vertical transport of particles? Does the consideration of a rough wall have an impact on the vertical transport of particles?

\section{Computational methodology and setup}\label{ch:setup}
\begin{table*}[ht]
\centering
\caption{Physical parameters of the simulation. Bulk Reynolds number $Re_b=U_bH/\nu$, cylinder Reynolds number $Re_D=U_bD_{cyl}/\nu$, friction Reynolds number $Re_\tau=u_\tau H_f/\nu$, particle Reynolds number $D^+$, density ratio $\rho_p/\rho_f$, cylinder diameter $D_{cyl}$, particle diameter $D_p$, Shields number $\theta$, Galileo number $Ga$, global solid volume fraction of all particles $\Phi_s$, global solid volume fraction of the mobile particles $\Phi_s^m$, Stokes number based on the near wall fluid time scale $St^+$ and Stokes number based on the bulk fluid time scale $St_b$. Note that the solid volume fraction of the mobile particles in the rough-wall case is identical to the one in the smooth-wall case. The values for the reference case are extracted from \cite{kidanemariam2013direct}. $H$ and $H_f$ denote the entire channel height and fluid height, respectively. For the smooth wall cases $H=H_f$, whereas in the rough wall case $H_f = H-0.8D_p$ was used.}
\label{Table1}
\begin{tabular}{c c c c c c c c c c c c c c}
\\ 
\hline
\hline
Case & $Re_b$	& $Re_D$ & $Re_\tau$ & $D^+$& $\rho_p/\rho_f$ & $D_{cyl}/H$ & $D_p/H$ & $\theta$ & $Ga$ & $\Phi_s$ & $\Phi_s^m$ & $St^+$ & $St_b$\\
\hline
ref    & 2870 & -    & 184.54 & 7.21  & 1.7 & -   & $10/256$ & 0.19 & 16.49 & $0.05\%$ & $0.05\%$ & 4.91  & 0.41\\
smooth & 3015 & 603  & 200    & 7.81  & 1.7 & 0.2 & $10/256$ & 0.2  & 16.49 & $0.05\%$ & $0.05\%$ & 5.76  & 0.44\\
rough  & 3015 & 603  & 210    & 8.45  & 1.7 & 0.2 & $10/256$ & 0.26 & 16.49 & $1.79\%$ & $0.05\%$ & 6.74  & 0.44\\
\hline
\hline
\\ 
\end{tabular}
\end{table*}

\begin{table*}[ht]
\centering
\caption{Numerical parameters of the simulation. Computational domain dimensions in the $i$th spatial direction $L_i$, number of grid points in the $i$th spatial direction $N_i$, grid resolution of the particles $D_p/\Delta x$, grid resolution of the cylinder $D_{cyl}/\Delta x$, grid spacing in wall units $\Delta x^+$, number of mobile particles $N_p^m$, number of fixed particles $N_p^f$, total observation time over which statistics are accumulated $T_{obs}$, number of three-dimensional flow fields that are saved during the simulation $N_{snapshots}^{3Dfields}$ and the number of particle snapshots which are saved during the simulation $N_{snapshots}^{part}$. The values for the reference case are extracted from \cite{kidanemariam2013direct}.}
\label{Table2}
\begin{tabular}{c c c c c c c c c c c}
\\ 
\hline
\hline
Case & $L_x \times L_y \times L_z$ & $N_x \times N_y \times N_z$ & $D_p/\Delta x$ & $D_{cyl}/\Delta x$ & $\Delta x^+$& $N_p^m$ & $N_p^f$ & $T_{obs}/T_b$ & $N_{snapshots}^{3Dfields}$ & $N_{snapshots}^{part}$\\
\hline
ref   & $12H \times H \times 3H$ & $3072 \times 256 \times 768$ & 10 & -    & 0.72 & 518 & 0     & 255  & 70  & 20314 \\
smooth & $18H \times H \times 2H$ & $4608 \times 256 \times 512$ & 10 & 51.2 & 0.78 & 577 & 0     & 1338 & 669 & 129800\\
rough  & $18H \times H \times 2H$ & $4608 \times 256 \times 512$ & 10 & 51.2 & 0.85 & 577 & 38404 & 220  & 111  & 22000 \\
\hline
\hline
\end{tabular}
\end{table*}
We consider open channel flow (driven by a spatially constant pressure gradient to maintain a time-constant flow rate) over a flat, smooth wall in horizontal arrangement. The flow field and the particle motion are taken as periodic in the streamwise and in the spanwise direction, while a free-slip condition is applied at the free surface. A rigid cylinder with diameter $D_{cyl}$ is mounted flush on the wall, spanning the entire flow depth up to the free surface. A dilute set of rigid, spherical, mobile particles with diameter $D_p$ (equivalent to approximately 8 wall units) are suspended in the flow. The length of the computational domain is chosen relatively large (equivalent to 90 cylinder diameters) such as to ensure a reasonable decay of the cylinder wake within the fundamental period. In the following $x$, $y$ and $z$ denote the streamwise, wall-normal and spanwise direction, respectively. Similarly, $u$, $v$ and $w$ denote the velocity in streamwise, wall-normal and spanwise direction, respectively. Time-averaged quantities are denoted by $\avg{}_t$, whereas time- and plane-averaged quantities are denoted by $\avg{}_{xzt}$.\\
Additionally, a second simulation over a rough wall consisting of fixed particles has been performed. The number of mobile particles was kept the same compared to the smooth wall case with the cylinder, such that the global solid volume fraction of the mobile particles is identical in both cases. One layer of fixed particles, all located at $y=D_p/2$, are arranged in a square arrangement of $418 \times 46$ (where particles overlapping with the cylinder have been removed). Another layer of fixed particles located at $y=-0.375D_p$ arranged staggered compared to the layer of fixed particles above the wall form spherical caps in the domain. This roughness arrangement was previously used by \cite{Chan-Braun2011}. Following \cite{Chan-Braun2011}, the virtual origin of the bottom wall was set to $y_0=0.8D_p$, such that the fluid height is $H_f=H-y_0$.
A summary of the salient physical and numerical parameters can be found in tables \ref{Table1} and \ref{Table2}, where the parameters from the reference case of \cite{kidanemariam2013direct} without a cylinder are repeated for convenience. Except for the presence of the cylinder and for the choice of the computational domain size, all parameter values are identical or very close to those in the study of \cite{kidanemariam2013direct}. 
Note that $u_\tau$ is computed based on the fraction of the driving pressure gradient that is balanced by the bottom wall contribution, denoted as $\avg{\Pi}_t^w$ with contributions from the flat wall and from the fixed particles in the rough wall case, such that
\begin{equation}
    u_\tau = \sqrt{\frac{-\avg{\Pi}_t^w H_f}{\rho_f}}
\end{equation}
\begin{figure*}
 \centering
 \begin{subfigure}[b]{0.6\textwidth}
     \centering
     \includegraphics[width=\textwidth]{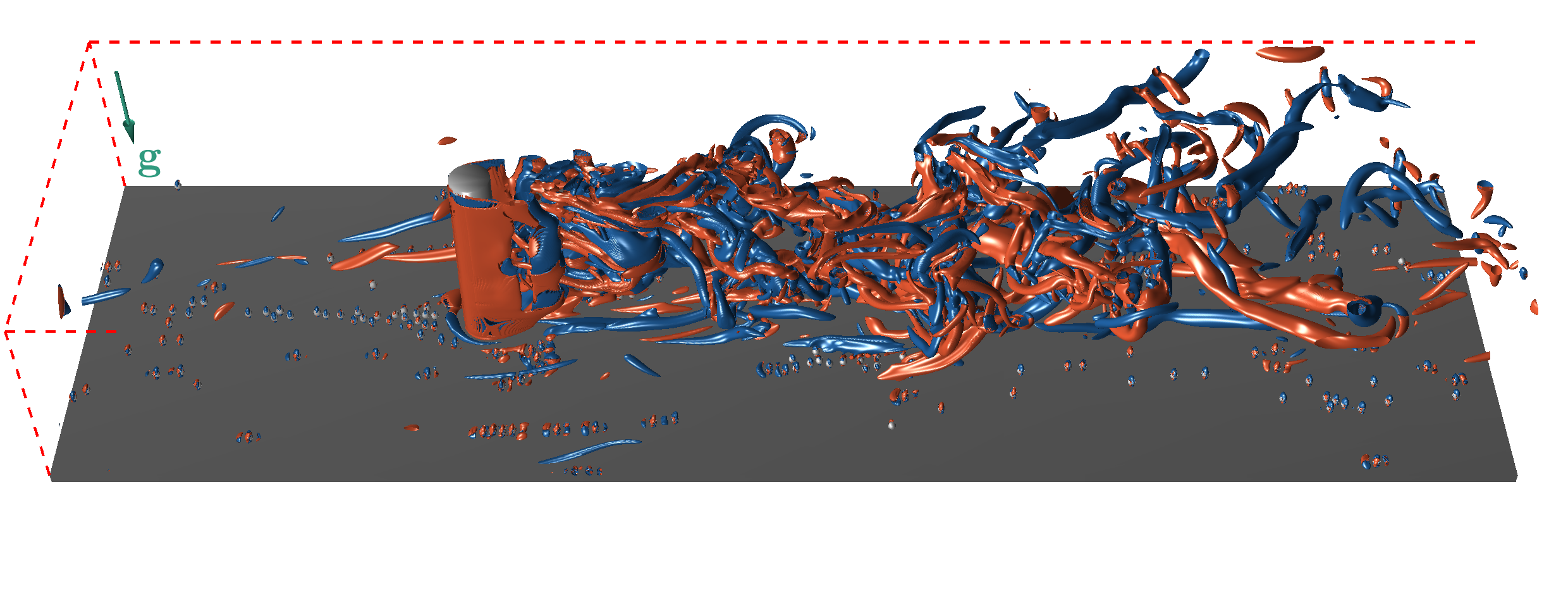}
     \caption{}
     \label{fig:Q_a}
 \end{subfigure}
 \hfill
 \begin{subfigure}[b]{0.3\textwidth}
     \centering
     \includegraphics[width=\textwidth]{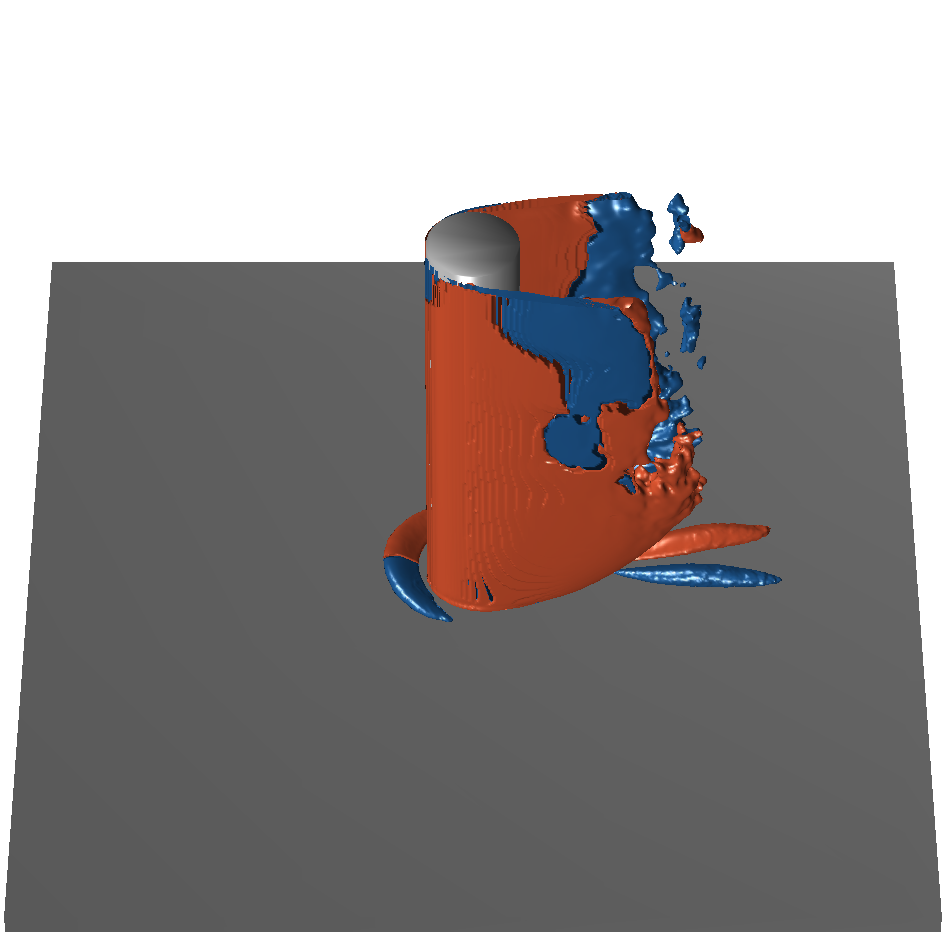}
     \caption{}
     \label{fig:Q_b}
 \end{subfigure}
 \caption{(a) Instantaneous and (b) time-averaged vortex structures in the cylinder wake, quantified by the second invariant of the velocity gradient tensor ($Q^+ = 0.05$) for the smooth-wall case. The colors show the direction of rotation determined from the sign of the streamwise vorticity $\omega_x$ (red: clockwise, blue: counter-clockwise). Note that the streamwise direction in (a) is only partly depicted for $x/D_{cyl} \in [-10,25]$. In (b) a close-up of the cylinder region is shown for $x/D_{cyl}\in [-5,5]$. The fluid flow is from left to right.}
 \label{fig:Q}
\end{figure*}
The numerical method is based on a fractional step formulation, a spatially second-order accurate (central) finite difference solver and a semi-implicit Runge-Kutta-3 / Crank Nicolson temporal scheme. The fluid-solid coupling both for the particles as well as for the cylinder uses the immersed boundary method as proposed by \cite{UHLMANN2005}. We have validated the cylinder implementation (without the presence of particles) through comparison with data obtained with the spectral solver Nek5000 \cite[]{nek5000-web}. Particle-particle contact as well as the contact between particles and the plane wall are treated with a short-range repulsion force which is active for gaps less than two grid spacings \cite[]{glowinski1999}. Contacts between particles and the cylinder are not included in the smooth-wall simulation. It has been verified that the number of particle-cylinder collisions is small and therefore negligible for the results.

\section{Results}
Figure \ref{fig:Q} shows both the instantaneous and time-averaged vortex structures around the cylinder. Additionally, the instantaneous particle locations are included in figure \ref{fig:Q_a}. The most intense vortical motion exists downstream of the cylinder. A horseshoe-shaped vortex system (HVS) can be seen upstream and on the sides of the cylinder both in the instantaneous and time-averaged visualization of the fluid field. The time-averaged flow field also includes two quasi-streamwise counter-rotating vortices in the cylinder wake near the wall separated by the symmetry plane. This pair is rotating such that high-speed fluid is brought down to the wall in the space between them.

\begin{figure}
 \centering
 \includegraphics{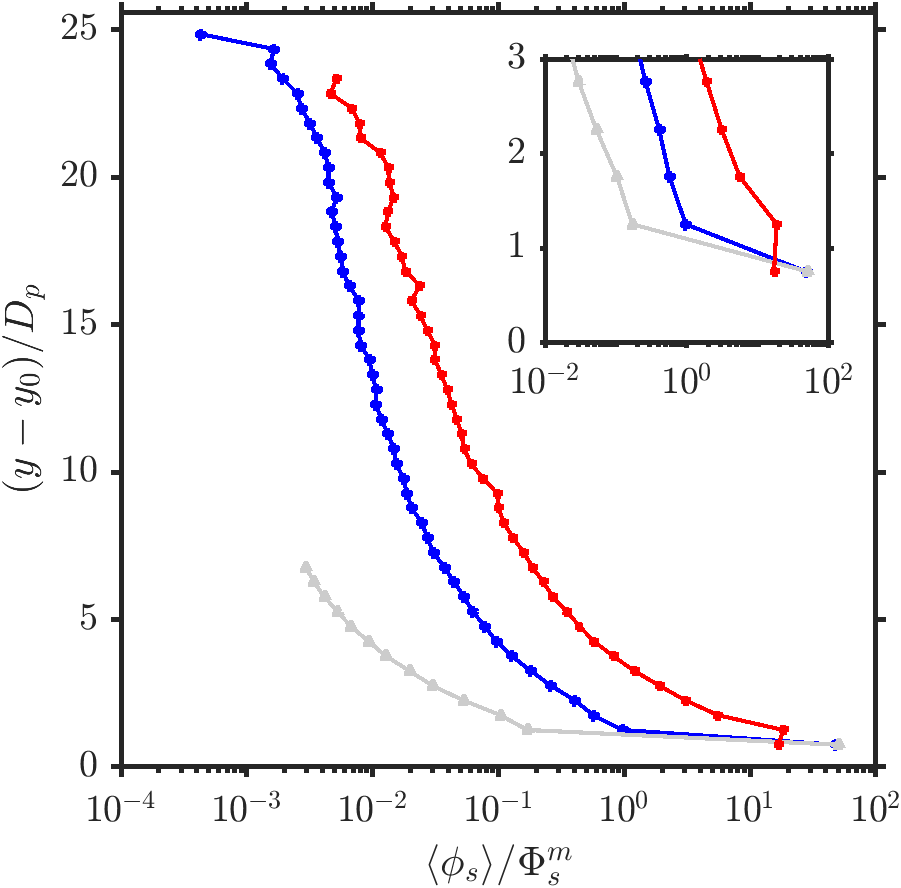}
 \caption{Time- and plane-averaged solid volume fraction computed in bins in $y$ direction ($\Delta h = D_p/2$) normalized by the global solid volume fraction of the mobile particles $\Phi_s^m$. The gray curve is the smooth-wall reference from \cite{kidanemariam2013direct} without the cylinder. The case with the smooth-wall including the cylinder is depicted in blue. The red curve corresponds to the case with a rough wall and the cylinder. The $y$-coordinate in the latter case was shifted by $y_0=0.8D_p$. The inset shows a close-up of the first six bins above the wall.}
 \label{fig:phi_s_over_y}
\end{figure}

Figure \ref{fig:phi_s_over_y} shows the wall-normal profiles of the time- and plane-averaged solid volume fraction. 
\begin{figure*}
 \centering
 \begin{subfigure}[b]{0.49\textwidth}
     \centering
     \includegraphics[width=\textwidth]{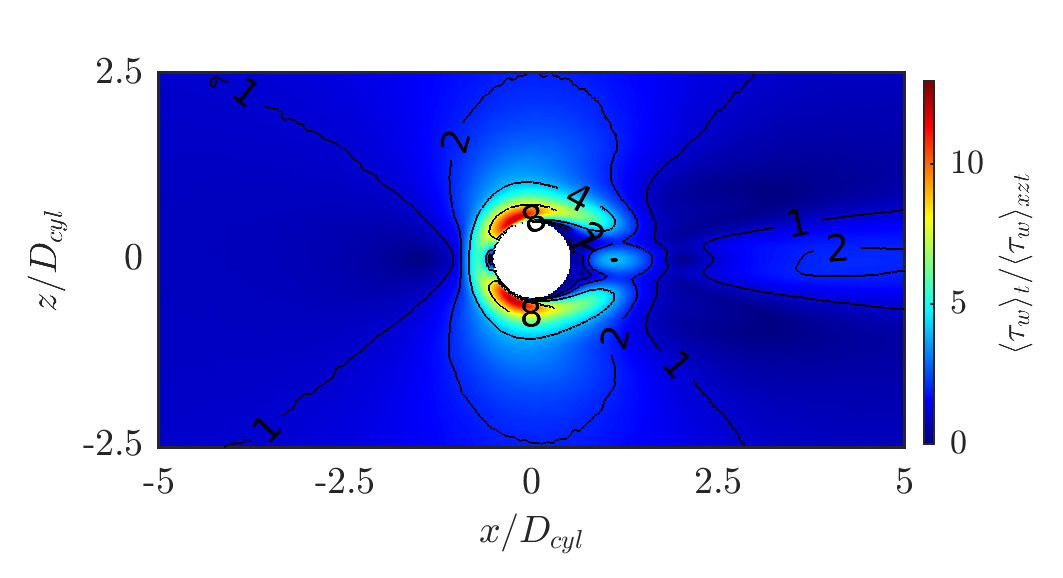}
     \caption{}
     \label{fig:tau_w_a}
 \end{subfigure}
 \hfill
 \begin{subfigure}[b]{0.49\textwidth}
     \centering
     \includegraphics[width=\textwidth]{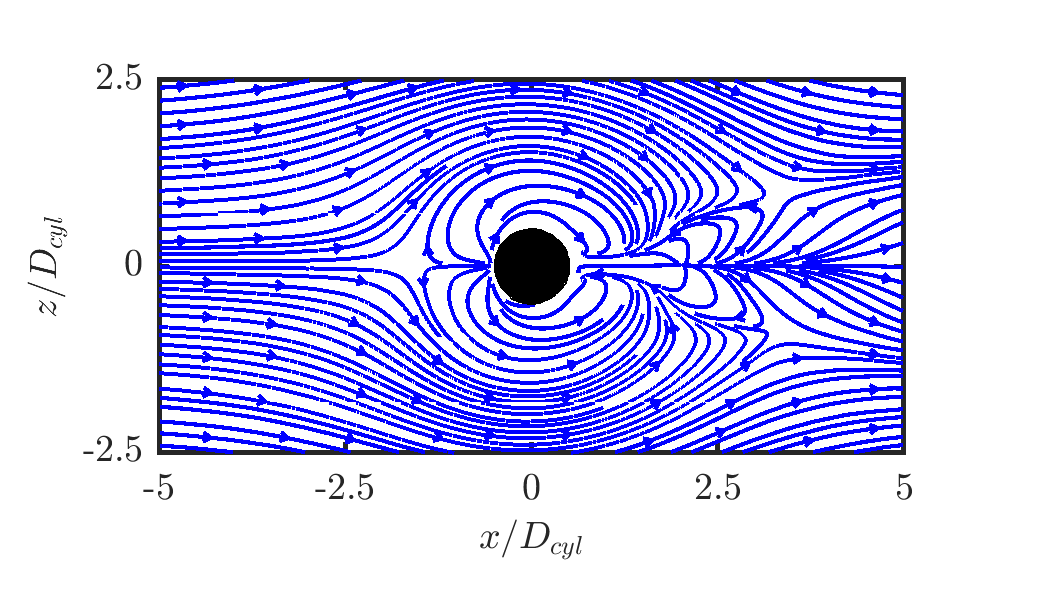}
     \caption{}
     \label{fig:tau_w_b}
 \end{subfigure}
 \caption{(a) Magnitude of the time-averaged wall-shear stress in the vicinity of the cylinder normalized by the time- and plane-averaged norm of the wall-shear stress for the smooth-wall case. (b) Tangential lines of the time-averaged wall-shear stress in the vicinity of the cylinder for the smooth-wall case.}
 \label{fig:tau_w}
\end{figure*}
The global solid volume fraction of the mobile particles has been used for normalization. Further details regarding the precise definition of the time- and plane-averaged solid volume fraction can be found in \cite{kidanemariam2013direct}. 
Due to gravity the majority of the particles is located in the vicinity of the lower wall, however, no particles are located inside the first bin which is a result of the applied bin width.
In the smooth wall case without the cylinder $99.15\%$ of the particles are located inside the second bin. In this case the data for $(y-y_0)/D_p > 7$ has been excluded due to the small amount of particles at these wall-normal distances.
In the smooth wall case with the cylinder $93.72\%$ of the particles are located in the second bin. The solid volume fraction in higher bins is significantly larger compared to the case without the cylinder. The flow modulation by the cylinder seems to increase the rate of particle entrainment.
For the rough wall case the locations of the bins are shifted by $y_0=0.8D_p$. Note that only the mobile particles are included in the computation of the time- and plane-averaged solid volume fraction as well as in the computation of the global solid volume fraction used for normalization. In the rough wall case $33.03\%$ and $35.87\%$ of the particles are located inside the second and third bin, respectively. The solid volume fraction for larger wall-normal distances is significantly higher compared to the smooth wall case with the cylinder. The roughness elements seem to add an additional mechanism to lift particles to higher wall-normal positions.
The lower number of particles in the second bin for the rough case compared to the smooth cases can be explained by the setup of the roughness elements, as introduced in chapter \ref{ch:setup}. Mobile particles which are located in the lowest possible position are classified into the second bin, whereas mobile particles which move across rows must inevitably move at least through the third bin. In the smooth wall cases, particles are able to slide along the smooth wall which is not possible anymore with the roughness elements. 
Note that the number of particle samples for the rough wall case is significantly lower compared to the smooth wall case due to the shorter runtime. The quality of the statistics especially away from the wall is therefore lower.

\begin{figure*}
 \centering
 \begin{subfigure}[b]{\textwidth}
    \centering
    \begin{tikzpicture}
        \draw (0, 0) node[inner sep=0] {\includegraphics[width=\textwidth]{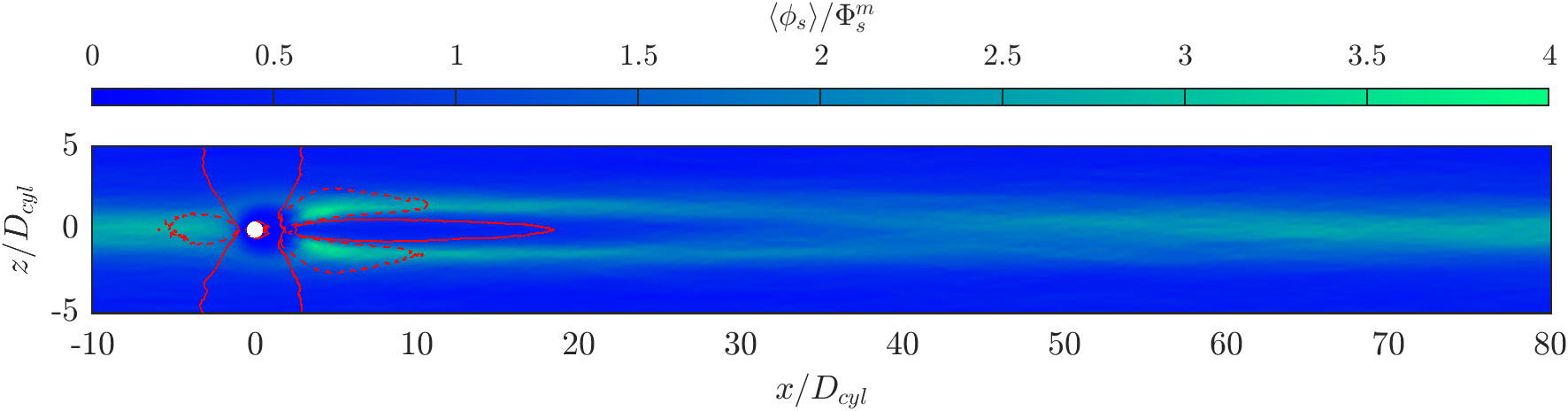}};
        \draw (-8, 2) node {(a)};
    \end{tikzpicture}
    \captionsetup{labelformat=empty}
    \caption{}
    \label{fig:phi_s_over_xz_a}
 \end{subfigure}
 \begin{subfigure}[b]{\textwidth}
     \centering
    \begin{tikzpicture}
        \draw (0, 0) node[inner sep=0] {\includegraphics[width=\textwidth]{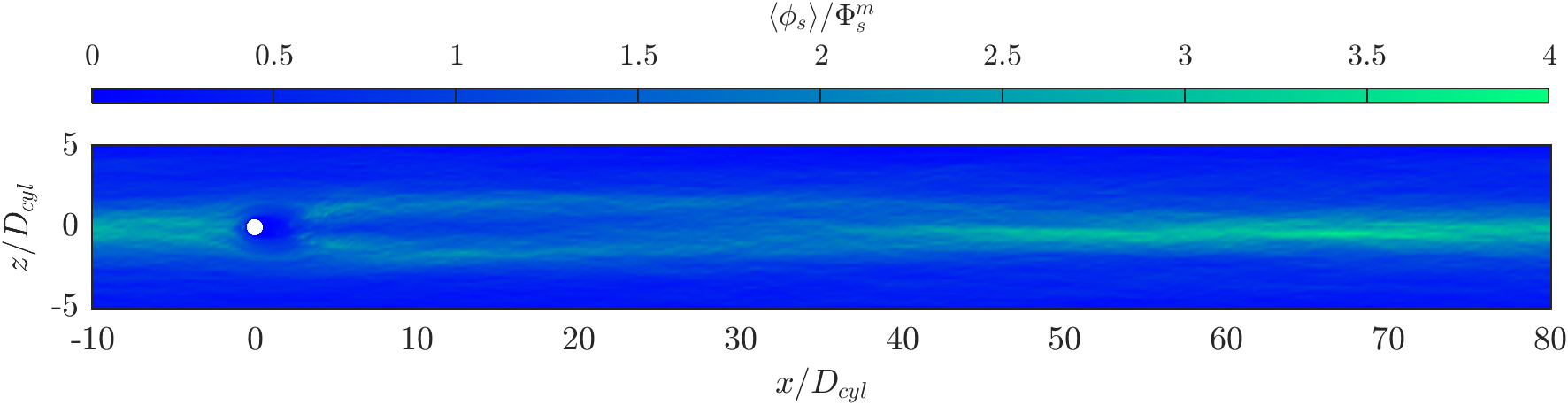}};
        \draw (-8, 2) node {(b)};
    \end{tikzpicture}
    \captionsetup{labelformat=empty}
    \caption{}
    \label{fig:phi_s_over_xz_b}
 \end{subfigure}
 \caption{(a) Time-averaged solid volume fraction computed in bins in $x$ and $z$ direction for the smooth-wall case with a respective bin width of $\Delta x_{bin}=\Delta z_{bin} = D_p$. The global solid volume fraction of the mobile particles was used for normalization. Note that the bins span the entire channel height $L_y$. (b) Same as in (a) for the rough wall case. The lines in (a) show the wall-shear stress fluctuation with respect to the plane-averaged value, defined as $(\avg{\tau_w}_t - \avg{\tau_w}_{xzt})/ \avg{\tau_w}_{xzt}$, for values of $0.2$ (solid line) and $-0.2$ (dashed line).}
 \label{fig:phi_s_over_xz}
\end{figure*}

Figure \ref{fig:tau_w} shows the distribution of the time-averaged wall shear stress in the vicinity of the cylinder for the smooth-wall case. 
Figure \ref{fig:tau_w_a} depicts the magnitude of the time-averaged wall shear stress vector $\avg{\tau_w}_t$, defined as
\begin{equation}
    \avg{\tau_w}_t = \sqrt{\left(\eval{\mu\frac{\partial \avg{u}_t}{\partial y}}_{y=0}\right)^2 + \left(\eval{\mu\frac{\partial \avg{w}_t}{\partial y}}_{y=0}\right)^2}
    \label{eq:tau_w}
\end{equation}
normalized by the time- and plane-averaged magnitude of the wall shear stress vector $\avg{\tau_w}_{xzt}$. Note that due to the flow modulation of the cylinder the time-averaged spanwise component of the wall-shear stress is non-zero. 
Around the cylinder a horseshoe-shaped region of increased wall shear stress can be observed. The maximal amplification of $\avg{\tau_w}_t/\avg{\tau_w}_{xzt} = 12.15$ can be found at an angle of $114^{\circ}$ to the streamwise direction at a distance of $r/D_{cyl}=0.59$ from the cylinder center axis. \cite{SCHANDERL2016} conducted LES in a comparable simulation setup at a much higher cylinder Reynolds number of $Re_D = 39000$. Their observed location of maximal amplification was located at $125^{\circ}$ with an amplification factor of $12$ (note that their normalization is based on the wall-shear stress simulation of a precursor simulation without the influence of the cylinder). In their work, the overall shape (including the maximum) seemed to have slightly shifted upstream, which might be a result of the increased Reynolds number. 
In the cylinder wake a region of increased wall-shear stress develops along the symmetry plane of the channel for $x/D_{cyl}>2.5$. This region is located between two regions of decreased wall-shear stress. This effect is later discussed in figure \ref{fig:phi_s_over_xz}.
Figure \ref{fig:tau_w_b} shows the tangential lines of the time-averaged wall-shear stress in the vicinity of the cylinder. This quantity is of relevance since the magnitude alone does not contain any directional information. In particular one can see the signature of the return flow on the upstream side of the cylinder axis which is a well-known feature in such junction flow \cite[]{simpson2001}.

Figure \ref{fig:phi_s_over_xz_a} shows the distribution of the time-averaged solid volume fraction in the $(x,z)$-plane for the smooth-wall case. 
The lines indicate values of $(\avg{\tau_w}_t - \avg{\tau_w}_{xzt})/ \avg{\tau_w}_{xzt}= \pm 0.2$ and highlight therefore regions of increased/decreased wall shear stress compared to the time- and plane-averaged value. 
In the close proximity of the cylinder the time-averaged solid volume fraction is low. This region matches the region of increased wall shear stress shown in figure \ref{fig:tau_w_a} well.
In the cylinder wake, two elongated regions of high time-averaged solid volume fraction separated by a central region of low time-averaged solid volume fraction exist up to roughly $x/D_{cyl} = 30$. This observation matches with the observed regions of increased/decreased time-averaged wall shear stress. The maximal value of the time-averaged solid volume fraction normalized by the global solid volume fraction is $3.83$.
Further downstream of the cylinder the two elongated regions of high time-averaged solid volume fraction merge and the majority of the particles is located in the vicinity of the symmetry plane. Due to the periodic setup most particles then approach the cylinder around the symmetry plane.
Figure \ref{fig:phi_s_over_xz_b} shows the corresponding data for the rough-wall case. In comparison to the smooth-wall case, the rough-wall case is not symmetrical about the symmetry plane, which is a result of the fewer samples over which the statistics are accumulated. Nevertheless, the same features as in the smooth-wall case are present: A region around the cylinder which is mostly avoided by particles and a similar pattern of streaks of particle agglomeration/depletion in the cylinder wake. It seems as if the region around the symmetry plane for $x/D_{cyl} < 30$ contains more particles compared to the smooth-wall case. This might however be a consequence of the insufficient convergence of the particle statistics for the rough-wall case.
%

\begin{figure*}
 \centering
 \begin{subfigure}[b]{0.49\textwidth}
     \centering
     \includegraphics[width=\textwidth]{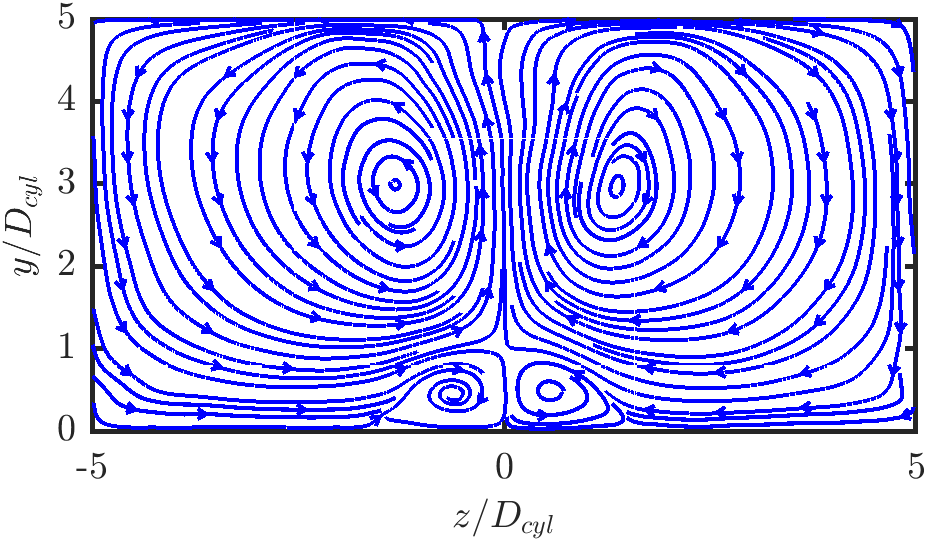}
     \caption{}
     \label{fig:vw_a}
 \end{subfigure}
 \hfill
 \begin{subfigure}[b]{0.49\textwidth}
     \centering
     \includegraphics[width=\textwidth]{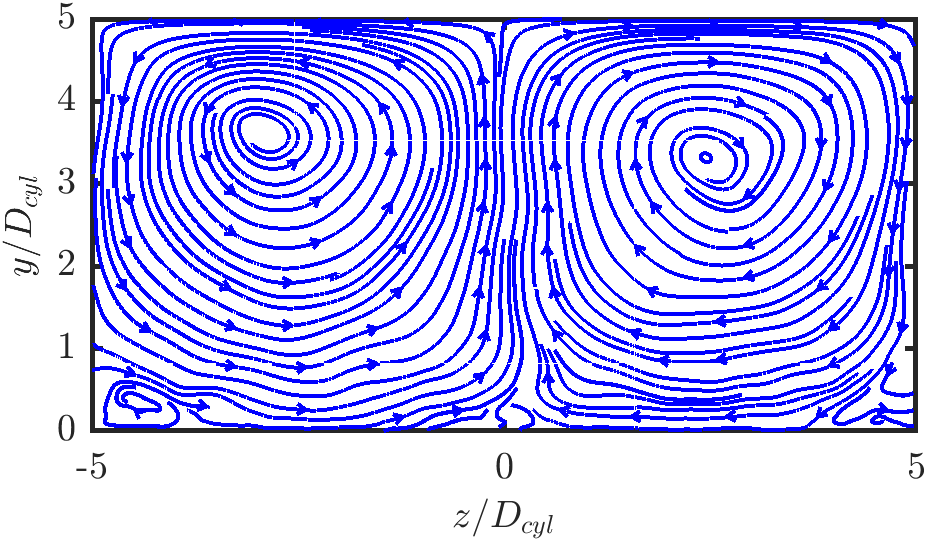}
     \caption{}
     \label{fig:vw_b}
 \end{subfigure}
 \caption{Time-averaged streamlines of $\avg{v}_t$ and $\avg{w}_t$ at (a) $x=7.6D_{cyl}$ and (b) $x=54.5D_{cyl}$, respectively, for the smooth-wall case.}
 \label{fig:vw}
\end{figure*}

Figure \ref{fig:vw} shows slices of the time-averaged streamlines of $\avg{v}_t$ and $\avg{w}_t$ at two $x$-positions for the smooth-wall case. In figure \ref{fig:vw_a} the secondary flow field at $x/D_{cyl}=7.6$ is depicted. Two large counter-rotating vortices in the outer region and two smaller near-wall vortices exist separated by the symmetry plane. Figure \ref{fig:vw_b} shows the secondary flow field for $x/D_{cyl}=54.4$, where the two smaller near-wall vortices have disappeared and only the two large counter-rotating vortices are present. The center of these vortices have shifted further away from the wall and symmetry plane. Other slices have been analyzed but are not shown here. It can be stated that the smaller near-wall vortices disappear roughly for $x/D_{cyl} = 30$.
This behavior explains why the two separated regions of high time-averaged solid volume fraction merge after roughly $x/D_{cyl} = 30$. For $x/D_{cyl} < 30$ the near-wall counter-rotating vortices tend to push near-wall particles away from the symmetry plane, whereas the larger vortices tend to push near-wall particles towards the symmetry plane. This ultimately results in the two elongated regions of high time-averaged solid-volume fraction separated by a region of low time-averaged solid-volume fraction.
For $x/D_{cyl} > 30$ the near-wall counter-rotating vortices have disappeared and only the large vortices further away from the wall still exist. Their action of pushing near-wall particles towards the symmetry plane is now not balanced anymore by the near-wall vortices, which then results in the observed streak of high time-averaged solid volume fraction around the symmetry plane.

\begin{figure*}
 \centering
 \includegraphics{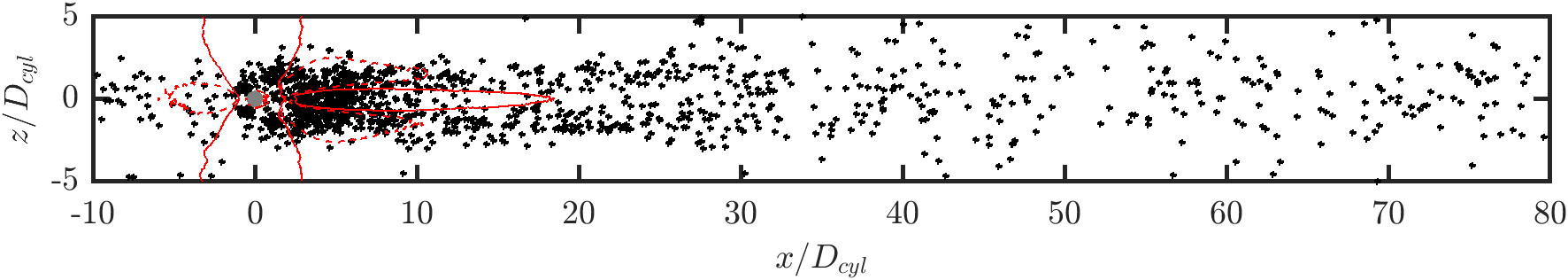}
 \caption{Locations of particle entrainment for particles exceeding the threshold height $y_{th}=5D_p$ for the smooth-wall case. The red lines show the wall-shear stress fluctuation with respect to the plane-averaged value, defined as $(\avg{\tau_w}_t - \avg{\tau_w}_{xzt})/ \avg{\tau_w}_{xzt}$, for values of $0.2$ (solid line) and $-0.2$ (dashed line).}
 \label{fig:lift-off}
\end{figure*}

Figure \ref{fig:lift-off} shows the detected locations of particle entrainment from the wall for particles which eventually exceed a threshold height $y_{th}=5D_p$ in the smooth-wall case. Different values for $y_{th}$ were examined and the sensitivity with respect to this parameter was found to be weak.
It can be seen that a region of increased entrainment rate exists downstream of the cylinder. These entrainments are probably caused by the increased vortical activity in this region, which was shown in figure \ref{fig:Q}. These entrainment events lead to a higher number of particles away from the wall for the smooth wall case with the cylinder compared to the case without the cylinder as shown in figure \ref{fig:phi_s_over_y}. The locations of particle entrainment in the wake of the cylinder for $x/D_{cyl} < 5$ seem to be independent of the distribution of the time-averaged wall shear stress. This might indicate that the particle entrainment in this region is not driven by the mean flow but rather by the action of instantaneous vortices, as depicted in figure \ref{fig:Q}. A thorough analysis of individual particle entrainments has not been conducted so far and is left for future studies. Further downstream until roughly $x/D_{cyl} < 25$, a central region with a reduced entrainment rate enclosed by two regions of higher entrainment rate can be observed. This pattern approximately matches the distribution of the time-averaged wall shear stress (and also the time-averaged solid volume fraction). 

\section{Conclusion}

In the present study we have investigated the motion of heavy spherical particles in horizontal open channel flow with a wall-mounted cylinder by means of PR-DNS both for a smooth wall and a rough wall. We have found that the presence of a cylinder significantly increases the amount of entrained particles compared to the case without a cylinder. In the rough-wall case with a cylinder the amount of entrained particles is even higher than for the smooth wall case with a cylinder.
Furthermore, a close link was found between the distribution of the time-averaged wall shear stress and regions of accumulation/depletion of particles. We have found that the close vicinity of the cylinder is typically avoided by particles and that streaks of preferred particle agglomeration/depletion exist in the cylinder wake, which are closely linked to the distribution of the time-averaged wall shear stress in the cylinder wake.
A region with an increased entrainment rate of particles for the smooth-wall case was identified around and directly downstream of the cylinder, such that the increased amount of particles at larger wall-normal distances can be attributed to the local flow modulation induced by the cylinder. Further work and additional samples are required in order to elucidate the detailed entrainment processes.


\bibliographystyle{tsfp}
\bibliography{tsfp}


\end{document}